# IMPROVING AODV PERFORMANCE USING DYNAMIC DENSITY DRIVEN ROUTE REQUEST FORWARDING

**Venetis Kanakaris, David Ndzi, Kyriakos Ovaliadis**

Department of Electronic and Computer Engineering,
University of Portsmouth, Anglesea Road, Portsmouth, PO1 3DJ, United Kingdom
www.port.ac.uk
`Email: {venetis.kanakaris,david.ndzi,kyriakos.ovaliadis}@port.ac.uk`

*ABSTRACT*

*Ad-hoc routing protocols use a number of algorithms for route discovery. Some use flooding in which a route request packet (RREQ) is broadcasted from a source node to other nodes in the network. This often leads to unnecessary retransmissions, causing congestion and packet collisions in the network, a phenomenon called a broadcast storm. This paper presents a RREQ message forwarding scheme for AODV that reduces routing overheads. This has been called AODV_EXT. Its performance is compared to that of AODV, DSDV, DSR and OLSR protocols. Simulation results show that AODV_EXT achieves 3% energy efficiency, 19.5% improvement in data throughput and 69.5% reduction in the number of dropped packets for a network of 50 nodes. Greater efficiency is achieved in high density network and marginal improvement in networks with a small number of nodes.*

## 1. INTRODUCTION

A combination of centralized and ad-hoc networks is envisaged to provide solutions for the provision of ubiquitous communication for a wide range of applications. Whilst centralized communication is well established, ad-hoc networking is seen as the way forward for self organizing and managing networks which eliminate meticulous and expensive planning, high cost, rigidity and vulnerability inherent in fixed centrally managed networks such as wired and wireless networks e.g. Mobile Telecommunications System (UMTS). Ad-hoc networks also hold great promise and applications in an extensive number of areas ranging from disaster management to environmental monitoring. Progress in ad-hoc networks is also facilitating the application of sensors for process automation in a variety of industries and is enabling progress in sensor fusion. Unpredictable events, e.g. earthquake often serve to illustrate the vulnerability of centrally managed networks and the importance of research and development in ad-hoc networks such as Mobile Ad-hoc Networks (MANET), for which centralized connectivity is not required. MANET is a wireless network that has mobile nodes with no fixed infrastructure. The main limitation of ad-hoc network systems is the availability of power and continuous reduction in the size of devices mean that power limitation cannot simply be ameliorated with large battery packs [1]. In addition to running the onboard electronics, power consumption is governed by the number of processes and overheads required to maintain connectivity.

A wide range of techniques have been proposed to address connectivity and power limitation issues in ad-hoc networks. These include hardware development, protocols, routing algorithms and battery technology or energy management systems [2]. Some researchers have proposed the development of hardware optimized for specific applications based on data rates [3]. More detailed studies of energy consumption by hardware have been carried out to evaluate energy consumption when transmitting, receiving, in sleep and idle modes. It has been proposed in [4][5] that energy management should be tailored to each application where the voltage, and hence processing speed and energy, can be reduced for non-time sensitive applications. This proposed technique benefits from the fact that the speed of microprocessors and energy consumption depend on the voltage that is applied to it. The common goal of energy management techniques proposed for ad-hoc networks is to preserve energy and maximize life span of the network. Other proposed methods are aimed at preventing network partitioning by managing energy consumptions of critical link nodes. In this paper we propose a modification of one of the most widely used protocols to improve the energy and data transmission efficiency of the network.

In general, ad-hoc wireless networks broadcast packets to the whole network as a means of transmitting information from one node to the other in the network [6]. Broadcasting in MANETs is not only a fundamental action for unicast routing protocols in mobile scenarios, but it is also an inextricable part of a number of multicast routing protocols. There is a diversity of geocast, unicast, and multicast protocols that use the broadcasting procedure in order to provide the significant function of control and route establishment. Broadcasting a packet to the entire network has extensive applications in mobile ad-hoc networks. Therefore, improving the broadcasting process will result in savings in several MANET applications.

Flooding is the simplest technique used by source nodes to broadcast packets to neighbouring nodes [7]. Each neighbour node receiving the packet for the first time rebroadcasts it ensuring outward propagation from the source until every node in the network has received and transmitted the broadcast packet exactly once.

Significant research activities have focused on reducing flooding in the network [7][8]. Any procedure that leads to a reduction in congestion saves energy and prolongs the life span of the network. In general, multi-hop transmissions are less energy efficient because of the startup energy consumption of the transceivers [9]. Therefore flooding which results in the reception and retransmission by multiple nodes in a network where path loss is not the dominant energy consumption element is energy inefficient. In high density networks allowing nodes to be turned off or enabling sleep mode and maximizing transmission range can increase energy efficiency. In [10] the concept of a minimum range routing where nodes within a specific range of a transmitting node are not allowed to retransmit a packet has been proposed. However, the proposed technique relies on nodes keeping an updated table of information about neighbouring nodes which can be time and energy inefficient in a high mobility environment. Another approach has been proposed in [11] where power consumption is distributed amongst the nodes by controlling the transmission and the reception powers. Using this technique, the amount of power consumed for sending one packet to any destination node is the same and determined for each node that is taking part in the routing process.

Broadcast protocols can be broadly divided into two main categories; deterministic and probabilistic. The probabilistic approach usually provide a simple solution in which every node that receives a broadcast packet has a fixed probability of forwarding the message [10]. But this approach does not guarantee full network coverage. On the other hand the deterministic approach can provide full coverage and can be further grouped into two categories, location information and neighbour set based.

In MANETs the routing task is delivered through network nodes which act as both routers and end points in the network. In order for a route to a specific destination node to be discovered, existing on-demand routing protocols use a simple flooding mechanism whereby a Route Request packet (RREQ) originating from a source node is broadcasted without exception to all nodes in the network [12]. This can lead to significant redundant retransmissions, causing high channel usage and packet collisions in the network.

In this paper there will be presented how the broadcasted routing messages react on the network performance. The protocols Ad-hoc On-Demand Distance Vector Routing (AODV), Dynamic Source

Routing (DSR), Destination Sequenced Distance Vector (DSDV) routing and Optimized Link State Routing (OLSR) have been studied thoroughly and their performances in simulated networks has been analysed. These protocols have been widely used and cited in literature [13].

*Types of MANET Routing Protocols*

One of the main tasks of routing protocols is to maintain routes inside MANET, since they do not use any access points to connect to other nodes in the network [14]. Routing protocols can be classified into three categories depending on their properties as follows:

- Centralized or distributed
- Static or adaptive
- Reactive or proactive

In centralized networks all route choices are made by a central node whilst in distributed routing networks the computation of routes is shared amongst the network nodes. In static routing, the route used by source and destination pairs is fixed regardless of traffic condition. It can only change according to the node needs or link failure. This kind of algorithm cannot get high throughput in cases of traffic conditions variety. In adaptive routing, the routes used between source and destination pairs may change in response to traffic condition e.g. congestion. There is another classification that the ad-hoc networks can be classified regarding the routing algorithms: proactive or reactive.

In proactive routing protocols, nodes maintain one or more routing tables about nodes in the network. These routing table information are updated either periodically or in response to a change in the network topology. The advantage of these protocols is that a source node does not need route-discovery procedures to find a route to a destination node. The disadvantage of proactive routing protocols is that because of it keeps an up-to-date routing table, creates essential messaging overhead, which consumes energy and bandwidth, and reduces throughput, especially when there is a large network with high node mobility. There are various types of table driven protocols which include: DSDV, OLSR, Wireless Routing Protocol (WRP) [15], Fish eye State Routing (FSR), Cluster Gateway Switch Routing (CGSR) protocol, and Topology Dissemination Based on Reverse Path Forwarding (TBRPF) protocol [16].

For reactive (on-demand) protocols there is an initialisation of a route discovery mechanism by the source node to find the route to the destination node when the source has data packets to send. When a route is found, route maintenance process is initiated to maintain this route until it is no longer required or the destination is not reachable. Reducing the message overhead is the advantage of these protocols. One of the drawbacks of these protocols is the delay in discovering a new route. Examples of reactive routing protocols include DSR, AODV and Temporally Ordered Routing Algorithm (TORA) [15].

This paper is organized as follows; Section 2 describes the protocols that will be evaluated in this paper. The routing procedure of AODV is described in more details and this is followed by the description of the proposed modification to AODV. Section 3 describes the simulated scenario, the settings, network configurations and the parameters that have been used to assess the performance of the protocols. The results of the simulation and discussions are provided in Section 4 which is followed by the conclusions.

## 2. ROUTING PROTOCOLS

For DSDV [17] protocol messages are exchanged between mobile nodes within range. Routing updates may be triggered or routine. Updates are initiated when routing information from one of the neighbours forces a change in the routing table. If there is a packet which the route to its destination is unknown, it is cached while routing queries are sent out. The data packets are stored temporarily until the destination node receive route-replies. The buffer has a size and time limit for caching packets beyond which packets are dropped.

All packets for which the route to their destinations is known are routed directly. In the event that a target is not found, the packets are forwarded to the default target which is the routing agent. The routing agent designates the next hop for the packet.

On DSR protocol the agent node controls every data packet regarding the information of source-route [18]. The packets are then forwarded as per the routing information. If there is no routing information in the packet, it gives the source route if route is known. When the destination is not known it caches the packet and sends out route queries. The routing query is initially sent to all nearby nodes and is always triggered by a data packet which has no route information regarding its destination. Route-replies are sent back if routing information to the destination is found.

AODV protocol is a mixture of both DSR and DSDV protocols [19]. It has the mechanism of route-discovery and route-maintenance of DSR. Moreover it keeps from DSDV the hop-by-hop routing sequence numbers and beacons. When a node needs to know a route to a specific destination, it creates a RREQ. The RREQ is forwarded by intermediate nodes which also create a reverse route from the destination. When the request reaches a node with a route to the destination node it also creates a Route Reply (RREP) which contains the number of hops that are required to reach the destination. All nodes that participate in forwarding this reply to the source node create a forward route to destination. This route is made not from complete route as in source routing but from every node which is a hop-by-hop state from source to destination.

OLSR [20] is a routing protocol where the nodes know all the available routes. As an optimized version of the pure link state protocol, the OLSR protocol floods the topological information to all active nodes in the network when the topology changes. A way to reduce the possible overhead in the network protocol is to use Multi-point Relays (MPR). The idea of MPR [21] is to diminish the number of duplicate retransmissions when a broadcast packet is forwarded. In this technique the number of retransmissions is restricted to a small set of neighbouring nodes, instead of using all the neighbours. This set is kept as small as possible by choosing the nodes which cover (in terms of one-hop radio range) the same network region as the complete set of neighbours. The OLSR routing protocol uses two types of control messages; Hello and Topology Control (TC). Hello messages are used in order to find out information regarding the link status and the host's neighbours. On the other hand TC messages are used for sending information about neighbours which includes the MPR selector list. The OLSR protocol has a disadvantage in that every host periodically sends the updated topology information to the entire network thus increasing the bandwidth usage. But this issue is solved by using the MPR, which forwards only the messages regarding the topology of the network.

## AODV Process

Normal RREQ and RREP processing mechanism of AODV is as follows:
- The source node S tries to send a packet to destination D.
- If S does not know the next hop for D, then it broadcasts a route request message.
- The RREQ message propagates in all directions to reach the destination D.
- The intermediate nodes that receive the RREQ message forward the packet to all its one hop adjacent nodes.
- If the destination, D, receives a RREQ message through a node N, then it sends a RREP to S by forwarding it to N since N may contain at least one routing table entry for S.
- On receiving the RREQ message through different nodes, the destination D will send the RREP message through different nodes and they may reach the source node through different possible paths.
- At the end, the source node S will have different possible resolved paths to select from based on defined criteria.

## Proposed modification of AODV RREQ mechanism

Standard AODV routing process broadcasts route request to all nodes. In the proposed scheme, a table of nodes in a given neighbourhood (one-hop nodes) is maintained. When a message is transmitted, only a subset of nodes in each neighbourhood is allowed to transmit. The number of selected nodes can be varied dynamically depending on the application and required quality of service. In this proposed scheme, the parameters that are used are defined in Table 1. Each node in the network will forward a route request message if and only if a condition based on its neighbourhood density at that instance is satisfied. The proposed scheme minimizes network congestion due to redundant transmission.

Table 1: Definition of AODV_EXT Parameters

| | |
|---|---|
| $n$ | The total nodes in the network |
| $F_i$ | Any node $F_i$, i = 1,2,…n that receives the RREQ message |
| $P_i$ | Packet forwarding probability derived from neighbour node count. |
| $\beta_i$ | The number of nodes neighbouring node $F_i$. |
| $d$ | Minimum number of neighbouring nodes - if the number of neighbours at a forwarding node, $F_i$, is less than or equal to $d$, then that node will forward the RREQ message to avoid path failure or network partitioning. |
| $C_f$ | It is a control factor which can be used to adjust the probability $P_i$ according to the application or average expected node density of the network, ($0 < C_f \leq 1$). |
| $R$ | Random number (between 0 and 100). This is used to generate varying conditions in the network. |

If the RREQ is received from an intermediate node then there will be at least one possible path which includes that node in its path list. Therefore, if only selected nodes are allowed to forward the RREQ packet, then only these nodes will be included in the path list. In this proposed scheme, the neighbourhood density of an intermediate node is considered as a criterion in RREQ forwarding decision at intermediate node. It means that if the number of nodes in the neighbourhood is high, then the probability of any node transmitting will decrease and hence reduces the transmission overhead. Random selection of nodes from the neighbourhood set increases the chance of full network coverage. Greater savings could be achieve by using a range dependent technique to select nodes for transmission but this can only be achieved at the cost of greater complexity.

---

**PROPOSED ALGORITHM**

Any node $F_i, i = 1,2,...,n$ receiving the RREQ message will process the packet as follows :

    For RREQ message originating from $S$ destined for node $D$ that is received by node $F_i$ process it if $F_i \neq S$ and $F_i \neq D$ (i.e. $F_i$ is an intermediate node) as follows:

    Node $F_i$ resolves its neighbourhood density $\beta_i$

    If $\beta_i \leq d$ then

        Forward the RREQ message

    Else

        Calculate message forwarding probability $P_i$ at node $F_i$

$$p_i = \frac{100}{\beta_i} * (d * C_f) \quad \text{for } 0 < C_f \leq 1$$

        If $R < p_i$ then

            Forward the RREQ message

        Else

            Ignore and Drop the RREQ message

        End

    End

## 3. SIMULATION AND METRICS

The simulation set up assumes the use of 802.11 standards based on the two path propagation model described by equation (1).

$$P_r = P_t G_t G_r \left(\frac{h_t h_r \lambda}{4\pi R^2}\right)^2 \quad (1)$$

Where $P_r$ and $P_t$ are the received and transmitted power; $G_r$ and $G_t$ are the gains of the receiving and transmitting antennas; $h_r$ and $h_t$ are the heights of the transmitting and receiving antennas; $\lambda$ is the wavelength of the signal and R is the distance between the transmitting and receiving nodes. This model assumes free-space. However, in reality the propagation conditions are usually more complex and often exhibit time and spatial variations resulting in shorter network life span than predicted by simulation. In addition, the simulation model takes an abstract view of the hardware power consumption by quantifying energy consumption based on the functions rather than hardware or changes in the propagation conditions. These assumptions are maintained in the simulation carried out in this paper because this can be accounted for by introducing a broad term for large scale variation. The energy model used can be described as follows [26]:

$$DecrTxEnergy = tx\_time * P\_tx \quad (2)$$

$$DecrRxEnergy = rx\_time * P\_rx \quad (3)$$

where DecrTxEnergy and DecrRxEnergy are the power consumed when transmitting and receiving, P_tx and P_rx are the power used when transmitting and receiving per unit time, respectively. The parameters tx_time and rx_time are the duration of transmission and reception, respectively. The lifetime of the network must take into account power consumed in three states: transmit, receive and idle. However, in this study power consumed in the idle state is not taken into account because the focus is on the effects of the protocols.

In the simulation configuration $d = 5$ was used. This defines a minimum condition which provides greater certainty of successfully packets routing but minimizes redundant transmissions. This is equivalent to minimum range routing but derived in terms of node density rather than distance.

Network Simulator 2 (NS2) has been used to evaluate the protocols [22]-[25]. The simulations were carried out to assess the performance of the routing protocols with network sizes of 10, 20, 30, 40 and 50 nodes with mobile node speeds between 1 m/s and 40m/s. For simplicity, in all cases the nodes send Constant Bit-rate (CBR) over User Datagram Protocol (UDP). The metrics that have been used to evaluate the performance of the network and protocols are the following [26]:

- *Number of packets dropped*: This is the number of data packets that are not successfully sent to its destination.
- *Consumed power*: The average consumed battery power.
- *Throughput*: This measure how fast the network can continuously send/receive data to the sink. Throughput is the number of packet received from the sink per millisecond.
- *MAC load*: This is the ratio of the number of MAC layer messages broadcasted from each node of the whole network to the number of data packets successfully delivered to all destination nodes. In other words, the MAC load is the average number of MAC messages generated for each data packet successfully delivered to the destination.
- *Control message overhead*: This control message overhead is the total routing control messages transmitted and received in the network.

The simulation configuration and specification are specified in Table 2.

## 4. RESULTS AND DISCUSSIONS

Figure 1 shows that AODV_EXT consumes less power than the other four protocols. Most importantly the power consumption of AODV_EXT based network improves in comparison to that of standard AODV as the number of nodes in the network increases beyond 30. The two protocols perform equally in small size networks. DSR protocol based networks consumes more energy compared to AODV and AODV_EXT but shows better performance when the number of nodes in the network is small (up to 30 nodes). On the contrary, the power consumption in networks using DSDV and OLSR rises steadily starting from fairly high levels. With increasing number of nodes, the energy depletion of OLSR based networks increases faster than those for the other protocols. OLSR protocol uses a mechanism that constantly updates information about nodes in the neighbourhood and therefore consumes more energy. As the number of nodes in the network increases, more updates are required and hence proactive protocols perform poorly, especially when the network is subject to changes e.g. in mobile environment.

Table 1: Simulation Parameters and Configuration

| **Routing Protocols** | AODV, AODV_EXT, OLSR, DSDV, DSR |
|---|---|
| **Topographical Area** | 800m x 800m |
| **Number of Nodes** | 10, 20, 30, 40, 50 |
| **Mobility** | 1m/s to 40m/s |
| **Channel type** | Wireless Channel |
| **Radio-propagation model** | TwoRayGround |
| **Network interface type** | WirelessPhy |
| **MAC type** | 802_11 |
| **Interface queue type** | DropTail/PriQueue |
| **Antenna model** | OmniAntenna |
| **Max packet in Queue** | 50 |
| **Transport /Traffic Type** | CBR over UDP |
| **TxPower of the nodes** | 0.1819 watts |
| **RxPower of the nodes** | 0.0501 watts |
| **IdlePower of the nodes** | 0.0350 watts |
| **Initial energy of the nodes** | 1000.0 Joules |

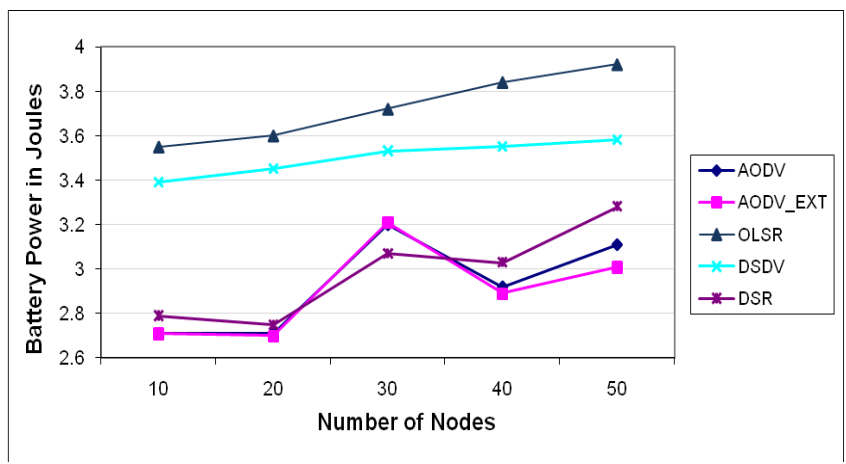

Figure 1: The Average Consumed Power

Figure 2 shows the performance of the protocols based on data throughput. It shows that AODV and DSR achieve comparable performance. However, AODV_EXT shows superior performance in larger networks. With AODV_EXT and AODV protocol, every node does not need to keep information regarding the route between two nodes. This reduces the amount of signaling required for route discovery and maintenance. OLSR and DSDV both show poor performances compared to the other three protocols. This is because both are proactive protocols and required table updates generate relatively high messaging overhead that can cause collision in large networks, especially in mobile networks, and reduces data rate performance of the network. However, these protocols are better suited to low data rate transmission because their self updating scheme ensures connectivity rather than the availability of bandwidth for application data.

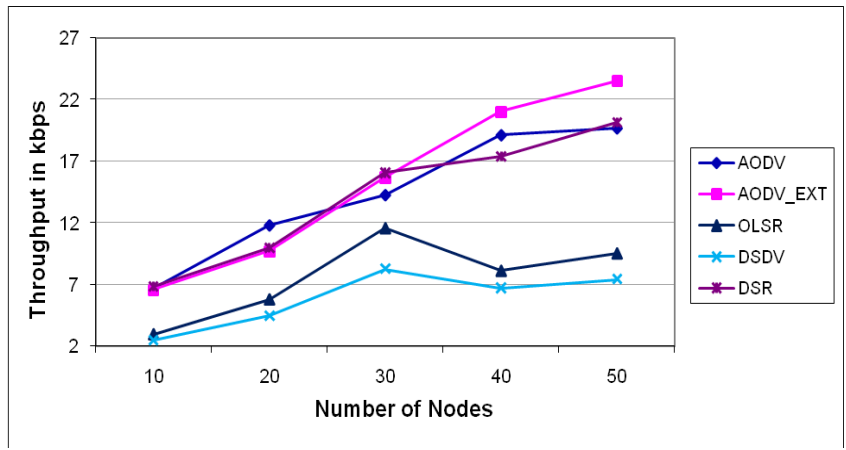

Figure 2: Comparison of data throughput for various network sizes (nodes)

Figure 3 shows MAC loading of the protocols. It shows that for large networks, a relatively high number of messages are generated by OLSR and DSR based networks. This increases sharply with the number of nodes in the network. DSDV and AODV exhibit only moderate increases. The figure shows that the use of density based scheme as applied in AODV_EXT significantly reduces the number of routing messages in the network. The proposed scheme reduces the amount of messages retransmitted and for a network of 50 nodes, the improvement is by a factor of two over standard AODV.

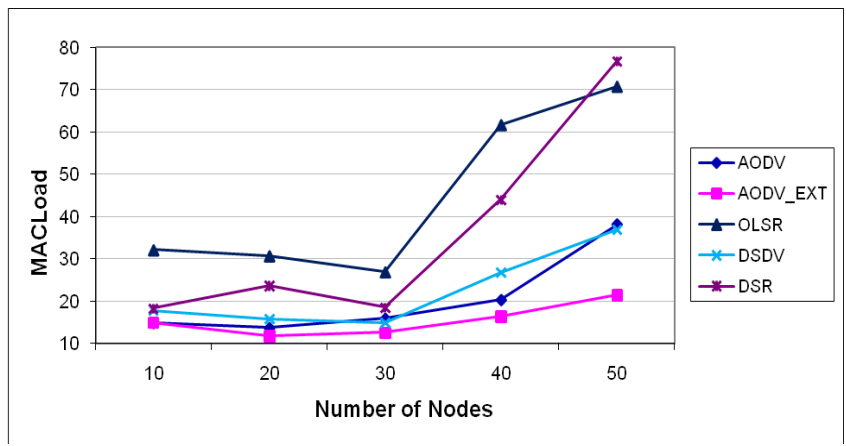

Figure 3 MAC Loading against the number of nodes

Figure 4 shows the routing control message overheads. In the case of AODV_EXT, it is lower than that of standard AODV protocol. DSR and DSDV show better results due to the fact that they transmit and receive the less number of control messages. OLSR protocol has the worst performance of all the protocols and this degrades significantly as the number of nodes exceeds 20.

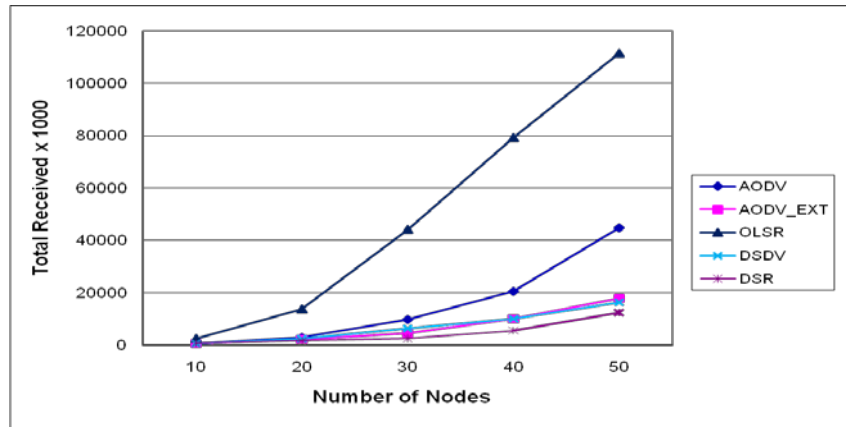

Figure 4 The Control Message Overhead

Table 2 provides details of the number of dropped packets for each protocol. DSDV drops the least number of packets with standard AODV dropping the most for a network of 50 nodes. However, compared to standard AODV, AODV_EXT performance is a factor 4 better.

Table 2: Number of dropped packets against network sizes

| Number of Nodes | DSDV | DSR | AODV | AODV_EXT | OLSR |
|---|---|---|---|---|---|
| 10 | 117 | 196 | 73 | 69 | 94 |
| 20 | 244 | 581 | 472 | 241 | 178 |
| 30 | 384 | 856 | 3011 | 766 | 473 |
| 40 | 911 | 4007 | 6537 | 1996 | 2978 |
| 50 | 1608 | 8880 | 17346 | 4056 | 7745 |

## 5. CONCLUSIONS

The assessment of four widely used protocols (AODV, DSDV, DSR and OLSR) has been presented in this paper. Their performances in different size networks and in mobile scenarios have been studied using simulations developed in Network Simulator 2 (NS2). AODV has been modified to use a probabilistic approach for transmitting route request messages. The modified version has been named AODV_EXT. Unlike in some probability based approaches where every node is assigned a fixed probability that does not guarantee full network coverage, the technique proposed in this paper combines concepts from maximum range node selection with node pruning to reduce redundant re-transmissions in route request but offer connectivity and better network coverage guarantees inherent in deterministic techniques.

The reduction in route request transmissions in a network using AODV_EXT has resulted in 3% energy efficiency savings, more than 60% reduction in the number of dropped packets because of reduced packet collision and increased data throughput. Moreover the results can be compared to that of Energy Reversed Ad-Hoc On-Demand Distance Vector (ER-AODV) routing protocol proposed by Khelifa and Maaza [27] which consumes up to 1.7 % more power than AODV_EXT. In addition, AODV_EXT improves the data throughput by more than 19% compared to the standard AODV and 10% more than ER-AODV. The results also show that proactive protocols, whilst they are more reliable in terms of connectivity, exhibit poor performance in large networks. Reactive protocols, on the other hand present better performances into large networks. Both classes of protocols perform poorly in large mobile networks due to large overheads associated with routing as the nodes move. A hybrid protocol such as AODV offer a compromise and the technique proposed in this paper to reduce redundant re-transmissions based on transmitting node neighbourhood density has produced very promising results when compared to standard protocols. This study has shown that fine tuning of protocols to suit specific applications or traffic scenarios to achieve optimum performance in ad-hoc networks will be essential.

## Authors

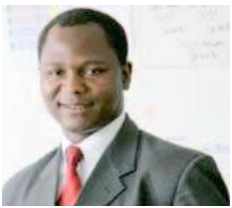
**Dr David L Ndzi** was born in Ndu, Cameroon. He graduated with Joint Honours degree in Mathematics and Electronics from Keele University in 1994 and with a PhD from University of Portsmouth in 1998. He worked as Research Associate from 1998 to 1999 and was appointed as lecturer in 1999. He is currently a Principal Lecturer in the Department of Electronic & Computer Engineering and also Faculty of Technology International Coordinator. His research focuses on wideband short and long range fixed and mobile wireless communications with emphasis on channel measurement, modelling and data rate optimisation in both planned and ad-hoc networks.

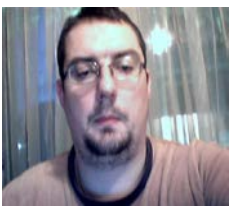
**Venetis Kanakaris** was born in Komotini, Greece. He received the B.Eng. degree in electronic engineering from Lamia Technical University, Greece, the B.Eng. degree in electrical engineering from Kavala Technical University, Greece, in 2004, and the M.Sc. degree in Telecommunications and Technologies from Technical University of Gabrovo, Bulgaria, in 2006. Since October 2009, he has been a research student in the Department of Electrical and Computer Engineering, University of Portsmouth, UK. His current research interests lie in the area of energy ad hoc routing protocols and low-power underwater wireless sensor protocol architectures.

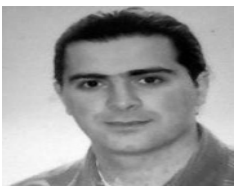
**Kyriakos Ovaliadis** was born in Kavala, Greece. He received the B.Eng. degree in electrical and electronic engineering from Coventry University, UK, in 1996, and the M.Sc. degree in electrical and electronic engineering from the University of Greenwich, UK, in 1998. Since February 2008, he has been a research student in the Department of Electrical and Computer Engineering, University of Portsmouth, UK. His current research interests lie in the area of low-power underwater wireless sensor protocol architectures and cluster based routing algorithms.